\documentclass[twocolumn]{aastex63}
\usepackage[caption=false]{subfig}
\usepackage{dcolumn}
\usepackage{tablefootnote}

\usepackage{lineno}
\shorttitle{Radio Structures in NGC~1068 with VLBA}
\shortauthors{Fischer et al.}

\begin{document}

\title{No Small Scale Radio Jets Here:\\[0.05cm] 
Multi-Epoch Observations of Radio Continuum Structures in NGC~1068 with the VLBA}

\correspondingauthor{Travis Fischer}
\email{tfischer@stsci.edu}

\author[0000-0002-3365-8875]{Travis C. Fischer}
\affiliation{AURA for ESA, Space Telescope Science Institute, Baltimore, MD, USA, 3700 San Martin Drive, Baltimore, MD 21218, USA}

\author[0000-0002-4146-1618]{Megan C. Johnson}
\affiliation{U.S. Naval Observatory, 3450 Massachusetts Ave NW, Washington, DC 20392-5420, USA}

\author[0000-0002-4902-8077]{Nathan J. Secrest}
\affiliation{U.S. Naval Observatory, 3450 Massachusetts Ave NW, Washington, DC 20392-5420, USA}

\author[0000-0002-6465-3639]{D. Michael Crenshaw}
\affil{Department of Physics and Astronomy,
Georgia State University,
25 Park Place, Suite 605,
Atlanta, GA 30303, USA}

\author[0000-0002-6928-9848]{Steven B. Kraemer}
\affil{Institute for Astrophysics and Computational Sciences,
Department of Physics,
The Catholic University of America,
Washington, DC 20064, USA}

\begin{abstract}

We present recent Very Long Baseline Array (VLBA) 5 GHz radio observations of the nearby, luminous Seyfert 2 galaxy NGC~1068 for comparison to similar VLBA observations made on 1997 April 26. By cross-correlating the positions of emitting regions across both epochs, we find that spatially-resolved extra-nuclear radio knots in this system have sub-relativistic transverse speeds (v $\lesssim$ 0.1c). We discuss sources of the observed knots and how the radio emission relates to additional phases of gas in the central $\sim$150 pcs of this system. We suggest that the most likely explanation for the observed emission is synchrotron radiation formed by shocked host media via interactions between AGN winds and the host environment.

\end{abstract}

\keywords{Active galaxies (17), Radio active galactic nuclei (2134), Radio astrometry (1337)}

\section{Introduction} \label{section: Introduction}

Active galactic nuclei (AGN) are the central regions of galaxies containing supermassive black holes that are actively accreting matter. This accretion generates intense radiation as matter falls toward the black hole and is rapidly heated up in the surrounding accretion disk. This process is complex and produces various forms of ``feedback'' as radiation drives material from the nuclear region and likely has some effect on how the host galaxy continues to form stars and evolve. The impact of these feedback processes continues to be studied, as direct connections between AGN feedback and galaxy evolution have not yet been fully established. 

Currently, two modes of feedback are utilized to describe interactions between the AGN and its host galaxy; `kinetic mode' and `radiative mode'. These individual modes are often used to describe the dominant form of feedback in distinctively different types of galaxies. Kinetic-mode feedback is often associated with radio galaxies and radio-loud quasars, typically massive elliptical galaxies exhibiting relativistic, collimated jets of synchrotron radio emission from pc to Mpc scales (\citealt{Har20} and references therein). These jets deposit their energy into the galactic and intergalactic environments, controlling the cooling of gas in galaxy halos and preventing star formation. Notably, low-power relativistic jets may also be capable of clearing out molecular gas in the central parts of galaxies \citep{Mor13,Muk18,Mur22}. Radiative-mode feedback occurs when energetic photons emitted by the AGN accretion disk couple with gas in the disk or host environment, creating winds that travel at speeds of hundreds to tens of thousands of km~s$^{-1}$, and are observed to propagate throughout the host on sub-pc to kpc scales. AGN winds likely impact the surrounding medium, either by removing significant amounts of gas and dust from the central regions of their host galaxies, which can presumably prevent star formation (e.g., \citealt{Sil98,Fab12}), or  instead compressing gas to promote star formation (e.g., \citealt{Sil13,Zub13}; also see the review by \citealt{Mor17}).

These modes have been defined over the last few decades and, individually, have been studied in great detail. However, as they are unlikely to be completely distinct, several studies have begun to examine the interactions between each mode and their overall impacts on their hosts. These interactions are often difficult to analyze in radio-loud AGN, as hosts either reside at distances where spatially resolved studies of AGN winds are difficult, such as in elliptical galaxies that lack significant amounts of gas and dust for AGN winds to interact with, or are inconveniently oriented such that evidence of AGN winds is largely extinguished by dust in the host galaxy (e.g., Centaurus A; \citealt{Ham15}). Alternatively, numerous radio-quiet AGN exhibit signatures that might indicate both modes of feedback: collimated, jet-like radio emission and high-velocity ionized-gas emission (i.e., \citealt{Fal98,Whi04,Gal06,Sto10,Pag12,Mak17,Fis19b,Jar19,Gir22}).

Observations of these targets indicate that the brightest radio structures are intertwined with ionized gas structures, known as the Narrow Line Region (NLR). In the Unified Model of AGN \citep{Urr95}, it is predicted that a radio jet will propagate perpendicular to the accretion disk. The jet is encompassed by an ionizing radiation field that exhibits a biconical shape which is constrained by the AGN torus. Therefore, to first order, the alignment between radio jets and gas ionized by the AGN is to be expected. However, the AGN-ionized NLR gas is also located on the surface of gas reservoirs in the host galaxies of these AGN \citep{Fis17,Shi19}. Therefore, radio emission appearing to be colocated with the NLR gas suggests that the observed jet not only aligns with the radiation field but also resides primarily embedded in gas reservoirs along the plane of the host galaxy. This consistent alignment has led us to suggest that the observed extended radio continuum emission in radio-quiet AGN is often not a jet pointed into the plane of the galaxy, but a by-product of radiatively driven winds originating at smaller radii, which extend to larger distances and drive into dense reservoirs to produce shocks \citep{Fis19b,Fis21}. These wind-driven shocks heat and compress the reservoir gas, producing significant amounts of cosmic rays that flow along enhanced magnetic field lines resulting from the compression, which in turn allow for the in situ production of synchrotron emission. In other words, we hypothesize that the radio emission observed radio-quiet AGN may be both a byproduct of radiative-mode feedback and an indicator of kinetic-mode feedback.

This hypothesis was supported in a recent study \citep{Fis21} of a volume-complete sample of 25 nearby AGN ($L_\mathrm{14-195~keV} > 10^{42}$~erg~s$^{-1}$, $D<40$~Mpc) selected from the Swift 105-month Burst Alert Telescope (BAT) catalog \citep{Oh18}, where we compared the peak-flux radio emission of the nuclear sources using the Jansky Very Large Array (VLA) and Very Long Baseline Array (VLBA) at the same sensitivity.

Assuming that the radio continuum in radio-quiet AGN can be attributed to a jet emanating from the nucleus, it is likely that a majority of the emission can be contained in a compact, or potentially unresolved, morphology when the jet is oriented along our line of sight and observed with the VLA. As such, the peak-flux measured over the nucleus with the VLA would be approximate to the total flux of the jet. When observed with the VLBA, a majority of this emission is then resolved out, creating a large VLA/VLBA flux ratio. Alternatively, if a jet is pointed into the plane of the sky, a small percentage of the total jet flux is unresolved over the nucleus when observed with the VLA. Therefore, when observed with the VLBA, a smaller portion of the emission from the entire jet is resolved out when compared to the peak-flux measured with the VLBA, creating a smaller VLA/VLBA flux ratio. From our study of the nearby BAT sample, we observed the opposite effect such that AGN with higher ratios of VLA to VLBA nuclear peak-fluxes tended to have extranuclear radio structures extended into the plane of the sky visible in the VLA observations.

If we instead assume that the observed radio continuum is not directly due to a jet but can instead be attributed to shocks through kinetic feedback processes, the ratio between VLA and VLBA fluxes would instead depend on the orientation of the AGN relative to its host galaxy. Radiative and kinetic feedback within the NLR and associated shocks would be generated where the host galaxy gas density is highest along host dust lanes often located in a disk. Therefore, targets that exhibit extended radio structures in the VLA would exhibit more interactions between the AGN and its host medium compared to targets with compact or unresolved VLA morphologies. We suggest that these interactions likely translate to parsec-scale levels, where the extranuclear host interactions may be unresolved into VLA peak-flux measurements and resolved out of VLBA peak-flux measurements.

\begin{figure*}
\centering
\includegraphics[width=1\linewidth]{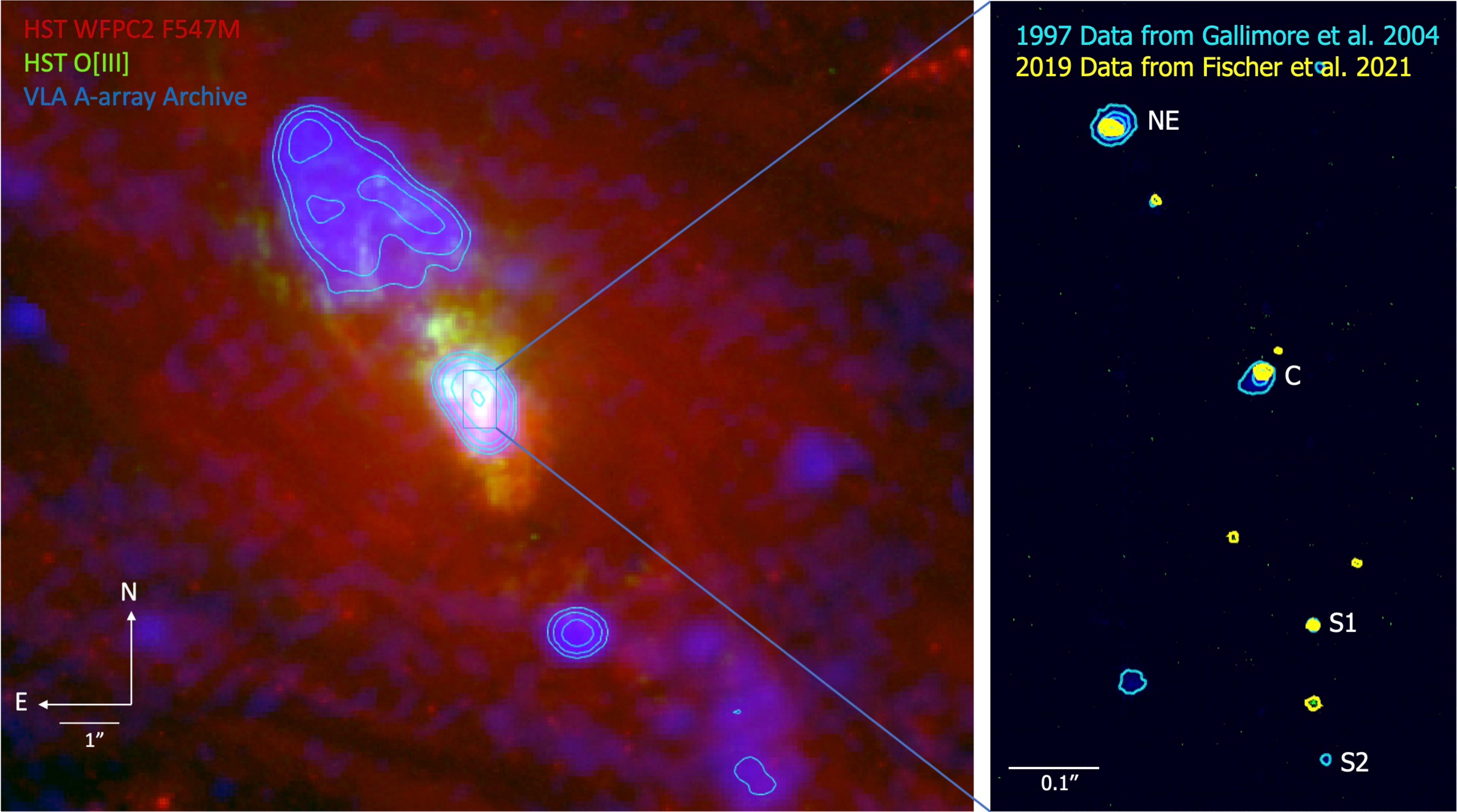}
\caption{Left: Composite image showing the central region of NGC~1068 where red is \emph{HST} broadband WFPC2 filter F547M, green is the continuum subtracted \emph{HST} [\ion{O}{3}], and blue is the VLA A-array 5 GHz continuum map. Right: Composite image showing the VLBA 6~GHz radio continuum data from \citet{Fis21} for NGC~1068 (yellow) with VLBA 5~GHz radio continuum data from \citet{Gal04} (cyan).}
\label{fig:vlba_comp}
\end{figure*}

This extranuclear contamination of VLA fluxes was best highlighted in NGC~1068. Figure~\ref{fig:vlba_comp} includes the employed VLA and VLBA observations from \cite{Fis21}, where the measured VLA peak flux is largely due to discrete extranuclear VLBA emission knot `C' north of the nucleus, without emission from the AGN itself (S1; \citealt{Gal04}). We return to the VLBA observations of this target in order to further study the characteristics of these extranuclear knots. Namely, as they are discrete and traceable outside the nucleus, we can study their change in projected distance over time to understand their relative transverse velocities. This will allow us to test whether these knots can be attributed to a relativistic jet. In this manuscript, we discuss the relative positions of several knots across a time span of 22 years, comparing observations from \citet{Fis21} to those described by \citet{Gal04}. In our analysis of NGC~1068, we use the Tully-Fisher distance to NGC 1068 of D = 10.1\,Mpc \citep{Tul09} and a scale of 48.9\,pc\,arcsec$^{-1}$.


\begingroup
\begin{deluxetable*}{l c c c c c c c c} \label{tab:distances}
\tablehead{\colhead{Knot}	& \multicolumn{2}{c}{Position Offset}    & \multicolumn{2}{c}{Projected Velocity}   & \colhead{$i$ at $V = 0.1c$\tablenotemark{a}} & \colhead{Velocity PA} & \colhead{S$_{\nu}$(1997)\tablenotemark{b}} & \colhead{S$_{\nu}$(2019)}\\
                            & \colhead{(mas)}  & \colhead{(pc)} & \colhead{(v/c)} & \colhead{(km s$^{-1}$})& \colhead{(degrees)}        & \colhead{(degrees)} & \colhead{(mJy)} & \colhead{(mJy)}}
\startdata
S1  & \nodata & \nodata & \nodata & \nodata & \nodata & \nodata & 9.1$\pm$0.8 & 2.6$\pm$0.2 \\
NE   &  3.29 & 0.16 & 0.024 & 7124 & 76.1  &  116.6 & 36.7$\pm$4.1 & 7.5$\pm$0.6 \\
C    &  6.09  & 0.30  & 0.044  & 13358  & 63.9 &  336.4 & 19.4$\pm$1.6 & 5.7$\pm$0.5 \\
\enddata

\caption{VLBA Extranuclear Projected Distance, Velocity, and Integrated Flux Measurements.}
\tablenotetext{a}{Inclination out of the plane of the sky required for a deprojected velocity of 0.1\,$c$}
\tablenotetext{b}{Integrated Flux measurements from \citet{Gal04}}

\end{deluxetable*}
\endgroup

\section{Data Analysis}

The VLBA data shown in the right panel of Figure~\ref{fig:vlba_comp} were observed 22 years apart.  We define Epoch~1 as the data from \citet{Gal04}, which we obtained from J.F.~Gallimore through private correspondence and Epoch~2 as the data from \citet{Fis21}.  For Epoch~1, we rebinned the image pixel scale to match the 0.8 mas pixels in Epoch~2. 

The nuclear emitting region in NGC~1068, labeled S1 in Figure~\ref{fig:vlba_comp}, as measured by \citet{Gal04}, exhibits water maser emission and likely marks the location of the central SMBH (\citealt{Gal01} and references therein). Previous analysis by \citet{Gal04} identified three additional extranuclear emitting regions, NE, C, and S2. Knots S1, NE, and C are clearly visible in both epochs. However, knot~S2 in Epoch~1 does not have a corresponding knot at its approximate location in Epoch~2. According to \citet{Gal04}, the integrated flux of source S2 was 4.1 $\pm$ 0.4 mJy at 5 GHz, thus, at the sensitivity of $\sim$22 $\mu$Jy/bm of the observations from \citet{Fis21} for NGC 1068, we would expect to have detected this source at a signal-to-noise greater than 180.  Thus, this source has either dropped in intensity such that it is below the sensitivity threshold presented in \citet{Fis21}, or, it has moved position. A potential corresponding emitter in Epoch~2 is north of the S2 position in Epoch~1, as shown in Figure~\ref{fig:vlba_comp}, as it is the brightest adjacent feature, roughly in the same radial path, and no emitting knots are observed south of the S2 Epoch~1 position in Epoch~2. However, as the relationship between these structures is unclear, we do not perform analyses for the distance between these knot positions.

\subsection{VLBA Correlations}

When the data from 1997 are directly compared to the data from 2019, none of the sources align astrometrically with one another and are offset by a few milli-arcseconds in the east-west direction. We investigated the possibility that the observed offset was caused by the phase calibrator used, but we have ruled this out. This is because we used the same calibrator, IERS 0237$-$027/J0239$-$0234, for both datasets in 1997 and 2019. This target is a defining source in the most recent International Celestial Reference Frame (ICRF3), which means it is one of the most astrometrically stable sources in the catalog. Additionally, the astrometric positional error of this calibrator, $\sigma_\alpha = \pm 2.09 \mu$s, $\sigma_\delta = \pm31.9 \mu$as, is lower than the observed offset between the two epochs. Through private communications with NRAO, we learned that the delay models for the VLBA were not well-constrained in 1997.  These delay models have been improved greatly over time and thus, the astrometry for the observations taken in 2019 is likely more robust.  Therefore, to begin our analysis of the features probed by the VLBA, we align the two datasets to the S1 structure as we believe this is the location of the AGN core and, therefore, likely static in position over time.  

In order to correct for relative offsets in nuclear position between epochs, we perform auto- and cross-correlation analyses on the position of knot S1. This analysis was performed using a large, $200\times200$~pixel ($0\farcs16\times0\farcs16$) base image and a smaller, $25\times25$~pixel ($0\farcs02\times0\farcs02$) subimage surrounding the S1 knot in both Epoch~1 and Epoch~2 using identical Epoch~2 ICRS coordinates. The subimage of S1 in Epoch~1 was auto-correlated against its base image to determine its Epoch~1 position in ICRS coordinates, defined as the auto-correlation distribution peak. We then cross-correlate the subimage of S1 knot from Epoch~1 against the base image from Epoch~2 and define the Epoch~2 position of the knot as the location of the cross-correlation distribution peak. The difference in relative positions between the auto- and cross-correlated distribution peaks then provides the projected distance offset between epochs. The S1 knot was aligned between epochs by altering the ICRS coordinates of the Epoch~1 data such that the distance between the Epoch~2 auto-correlation peak and cross-correlation peak is zero.

With the S1 knot from both epochs aligned, we repeated the cross-correlation measurements for the C and NE knots to measure their differences in position over time. We provide this difference for each knot in Table~\ref{tab:distances}. From our measurements of the aligned images, we find angular distances between correlation peaks for the C and NE knots of 6.09 and 3.29\ mas, respectively. At the given distance for NGC~1068, these differences correspond to a few tenths of a parsec. Knowing the difference in time between observation epochs, $t = 6.93\times10^{8}$\,s, we can calculate the projected velocities of knots C and NE to be $\approx$ 13300 and 7100 km s$^{-1}$, respectively, a few percent of the speed of light. Using the position offset between the two epochs, we can also measure the position angle of the velocity vector in the plane of the sky, with Knot~C traveling northwest at a position angle of $336\fdg4$ east of north and Knot~NE traveling southeast at a position angle of $116\fdg6$ east of north. We note that these vectors are roughly perpendicular to the position angle of the overall radio structure in this system, PA~$\approx30\arcdeg$, as illustrated in Figure~\ref{fig:vlba_comp}.

\section{Results and Discussion} \label{section: Results and Discussion}

At projected values between approximately $7,000 - 14,000$~km~s$^{-1}$, the velocities observed for knots NE and C are $<10\%$ the speed of light. The magnitude of these velocities is relatively unusual in terms of AGN feedback processes. As described above, jets in kinetic-mode feedback are relativistic, traveling at a large fraction of the speed of light. This is highlighted in observations by the MOJAVE program \citep{Lis16}, which illustrate the relativistic motion of radio-emitting knots in radio-loud quasars over several decades. This is also highlighted in theoretical models \citep{Muk18} where jets interacting with and affecting dense, gaseous disc material also require highly relativistic ($v/c > 0.4$) jets within the inner $\sim200$~pc to couple with the gas and launch outflows. On the other hand, optical winds in radiative-mode feedback are much slower (V$_{max} <$ 2000 km s$^{-1}$; \citealt{Fis13,Fis18}), largely restricted by the increasing enclosed mass of the host galaxy as a function of outflow radius. As our measurements are projected values, we also calculate the inclination of the velocity vector that is required for knots NE and C to reach 0.1c, provided in Table~\ref{tab:distances}, for a general understanding on the likelihood that we are observing projected, relativistic motion. Knot~C requires an inclination of $\sim64\arcdeg$ from the plane of the sky to reach 0.1c and Knot~NE would require a larger inclination of $\sim76\arcdeg$. Inclinations required to reach velocities of $\sim$0.2c for either knot are greater still, between $78\arcdeg - 84\arcdeg$ away from the plane of the sky, which suggests that the knots would need trajectories near parallel with our line of sight and perpendicular to the radiative outflows  to be traveling at a significant fraction of the speed of light.

\begin{figure}
\centering
\includegraphics[width=0.95\linewidth]{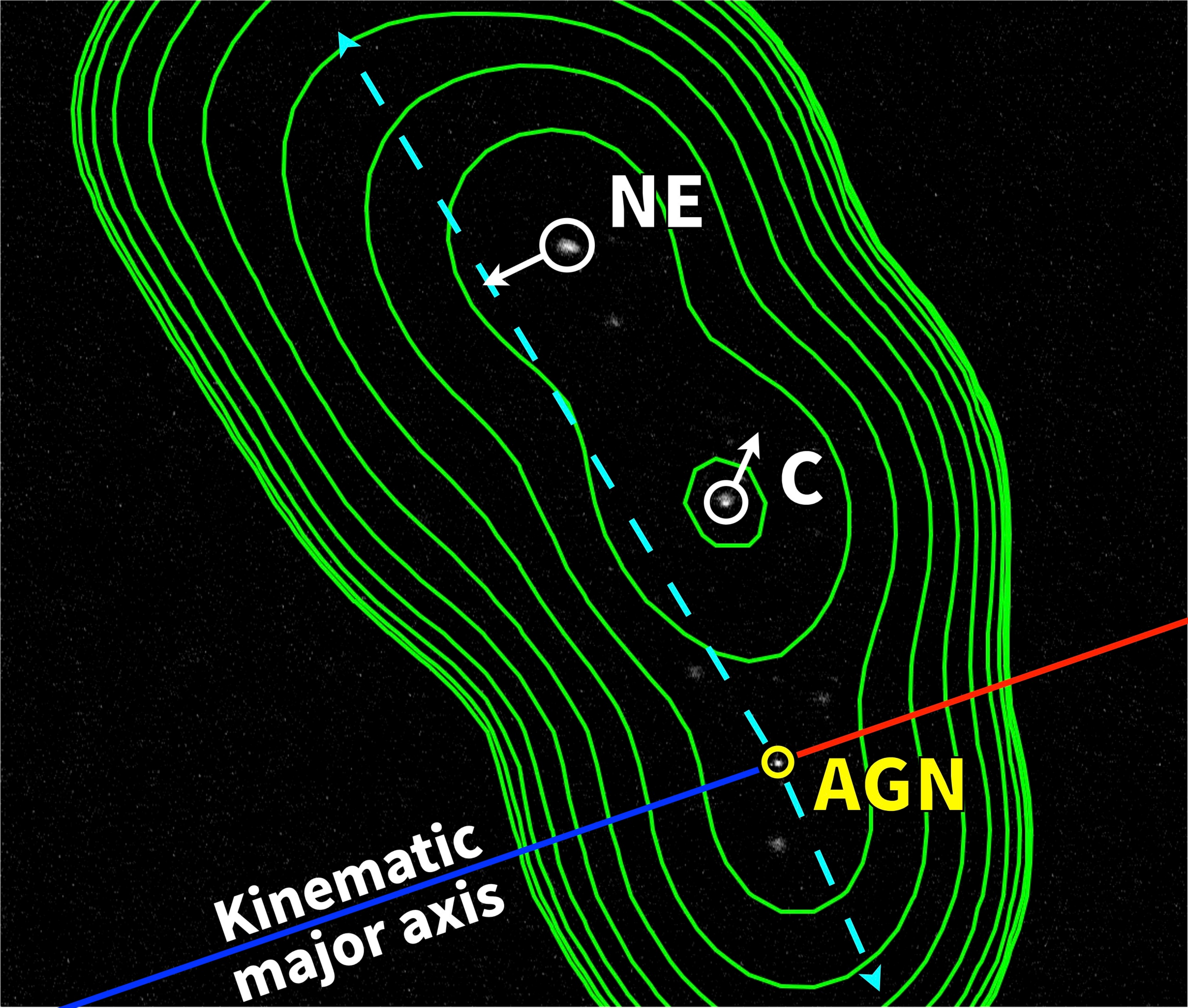}\\
\caption{Epoch~2 VLBA 6~GHz radio continuum data with archival VLA A-array 8.49~GHz continuum imaging contours (green) overlaid. Knots NE and C in Epoch~2 are circled in white with arrows pointing in the direction of their measured velocity vectors (not to scale). The central AGN is circled in yellow. Cyan, dashed lines point to the apex positions of the extended radio sources shown in Figure~\ref{fig:vlba_comp}. The kinematic major axis of the rotating molecular gas, per \cite{Gar19}, is shown as blue and red lines, with rotation traveling counterclockwise on the plane of the sky.}
\label{fig:vectors}
\end{figure}

Further, knots NE and C do not exhibit velocity vectors traveling radially away from the AGN, illustrated in Figure~\ref{fig:vectors}, as one would expect through AGN feedback processes. Paths to the most extended radio emission to the north and south of the AGN, visible in Figure~\ref{fig:vlba_comp}, are also shown in Figure~\ref{fig:vectors} which are approximately in the same direction as the normal of the radiation bicone (PA~$\sim 30\arcdeg$; \citealt{Das06}). We measure the position angles of the knot velocity vectors, provided as angles east of north, to be perpendicular to these axes. 

If we define a radio jet to exhibit relativistic velocities, we set 0.1$c$ as a lower limit for relativistic motion, as this velocity is where one is typically required to begin accounting for relativistic effects. Comparing the observed velocities and trajectories to this definition, we do not find the observed radio emission to exhibit the relativistic velocities present in radio jets. As described earlier, multi-epoch observations of jetted structures in radio-loud AGN have shown the velocities to move at relativistic velocities in a radial motion away from the central AGN. Jet kinematics can be more complicated at smaller scales. M87, a nearby radio-loud AGN, exhibits an intricate, polarized double-helix morphology in its radio jet \citep{Pas21}. One could posit that the velocity vectors observed in NGC~1068 are due to a similar scenario; however several clues in our observations suggest this helical, spiral pattern is not occurring. Immediately, the observed motions are not in a tight and orderly pattern as seen in jets. Also, we know that the observed knots are not individual point sources or narrow streams that could be traveling along magnetic field lines. As traced by VLBA, MERLIN, and VLA radio observatories at decreasing resolution (Figures \ref{fig:vlba_comp}, \ref{fig:vectors}; see also \citealt{Gal04}), the observed extranuclear knots are instead peaks of much larger structures. 

\subsection{Scenarios to Produce Observed Knot Velocities}

\begin{figure*}
\centering
\includegraphics[width=0.90\linewidth]{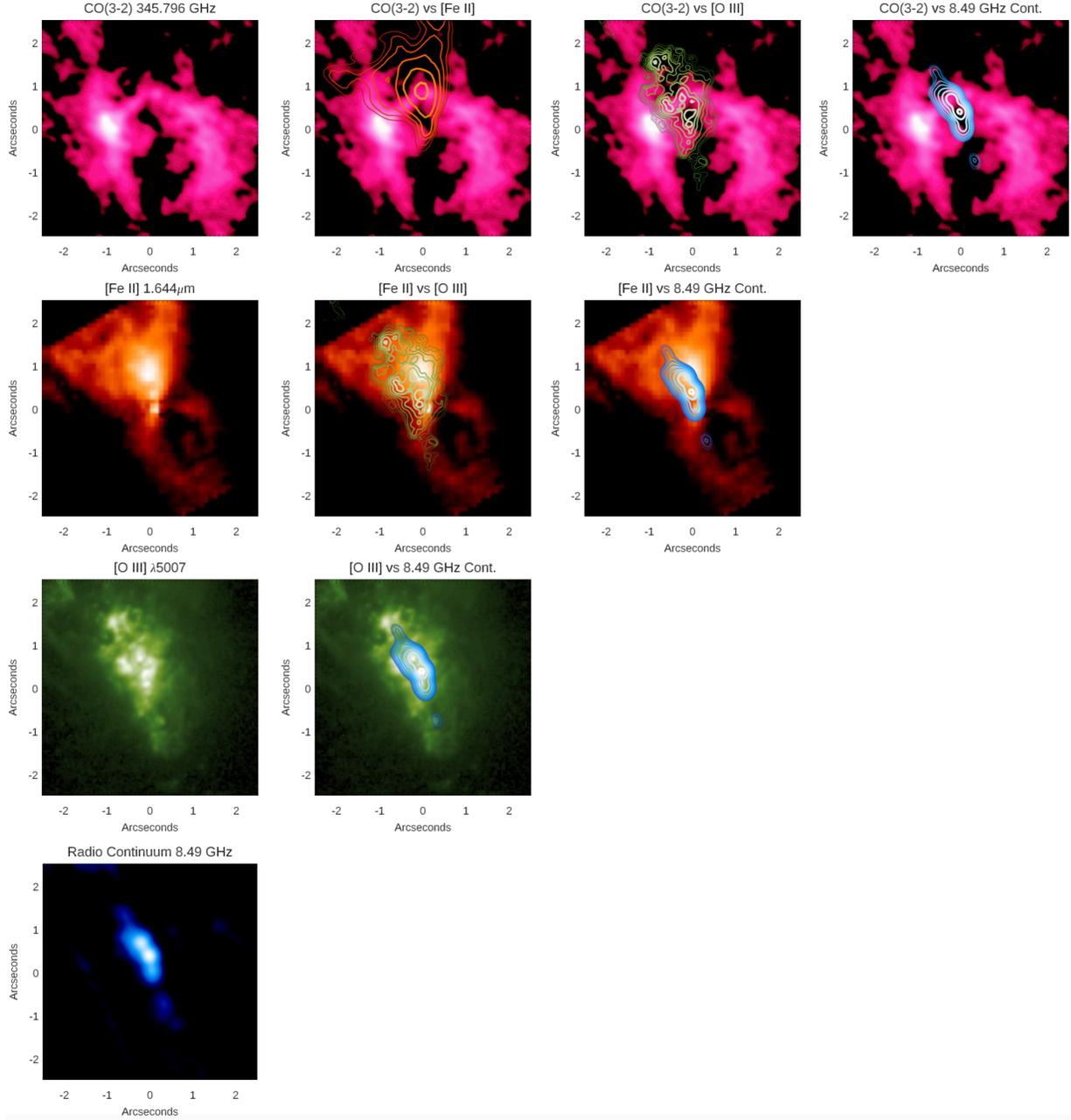}\\
\vspace{-2.0cm}
\caption{Multi-band imaging of the nuclear region of NGC~1068, centered on the approximate location of the AGN. First row: Continuum-subtracted ALMA Band 7 observations of CO(3-2) 345.796 GHz emission with contoured comparisons of [\ion{Fe}{2}] 1.644$\mu$m (second column; red), [\ion{O}{3}] $\lambda$5007 (third column; green), and 8.49~GHz radio continuum (fourth column; blue) features. Second row: Continuum-subtracted Gemini NIFS IFU imaging of [\ion{Fe}{2}] 1.644$\mu$m emission with contoured comparisons of [\ion{O}{3}] $\lambda$5007 (second column; green), and 8.49~GHz radio continuum (third column; blue) features. Third row: Continuum-subtracted {\it HST} WFPC2 imaging of [\ion{O}{3}] $\lambda$5007 emission with a comparison to 8.49~GHz radio continuum (second column; blue). Bottom row: VLA A-Array imaging of 8.49 GHz radio continuum. Outer contours of [\ion{Fe}{2}], [\ion{O}{3}], and radio continuum imaging represent S/N of 12$\sigma$, 100$\sigma$, and 10$\sigma$, respectively. North is up and east is to the left.}
\label{fig:contours}
\end{figure*}

Assuming the production of radio emission is due to shocks, there are several possible candidates for the responsible radiative-mode winds. First, the [\ion{O}{3}] outflows themselves. Kinematic analyses \citep{Das06, Ven21}, show that the FWHMs near the radio knots are large, greater than 1000 km s$^{-1}$, but there is no direct evidence for [\ion{O}{3}] enhancement specifically at the knot locations, per Figure \ref{fig:contours}, described below. Alternatively, the shocks could be created via X-ray winds. Work by \cite{Kra20} and \cite{Tri21} revealed X-ray emission-line regions with masses and mass-outflow rates comparable to, or greatly exceeding, those of the optical emission-line gas, providing an order of magnitude larger kinetic power. Finally, ultra-fast outflows (UFOs; \citealt{Tom10}) reach velocities $>$ 0.2\,$c$, which should have sufficient energy to cause strong shocks \citep{Mak23}.

Although the progenitors of these shocks potentially have relativistic velocities, we suggest that the observed difference in position between epochs does not require the individual knots to travel $\sim$10000 km s$^{-1}$ between the observed epochs. Instead, as the relationship between the observed radio-continuum emission and host medium is in a turbulent and dynamic environment, the impact point at which AGN feedback creates these shocks is likely what changes over time. In this scenario, the interaction between AGN winds and the surrounding environment changes the morphology of the dense molecular gas reservoir observed in each epoch. As gas in this reservoir is ionized and driven away, the remaining gas is sculpted to possess a new morphology. Winds then impact and shock different surfaces of the dense host medium due to these ablative processes to create the apparent motion of the radio emission.

\subsection{Multiwavelength Nuclear Imaging}

To illustrate how radio emission forms within the host environment, we compare its structure to adjacent structures in other wavebands.  Figure~\ref{fig:contours} compares the morphologies and locations of different forms of emission surrounding the nucleus of NGC~1068 across several wavebands and observatories; ALMA Band 7 observations of CO($3-2$) \citep{Gar19}, Gemini/NIFS H-band observations of [\ion{Fe}{2}] 1.6$\mu$m \citep{Rif14,Bar14}, HST/WFPC2 narrow-band imaging of [\ion{O}{3}] $\lambda$5007 \citep{Sch03,Fis18}, and VLA A-array observations of 8.49 GHz radio continuum (observed 8 September 1999, NRAO/VLA Archive Survey), which provides a large-scale comparison to our VLBA observations, as shown in Figure~\ref{fig:vectors}. ALMA and Gemini observations were aligned using the location of the AGN provided in previous analyses \citep{Rif14,Gar19}. HST [\ion{O}{3}] imaging alignment is described in Appendix \ref{section: alignment}.

We have previously analyzed connections between ionized gas and molecular gas \citep{Fis17} and shown that molecular gas is the reservoir from which spatially-resolved, ionized-gas structures are formed. We observe this relationship again in NGC~1068, as [\ion{O}{3}] is largely observed north of the AGN, and peaks of [\ion{O}{3}] emission-line knots can be mapped to CO reservoir walls facing the NLR axis east and west of the nucleus. Kinematic measurements in these regions suggest the gas is highly turbulent, with complex emission-line profiles and large mass outflows \citep{Das06,Rev21,Ven21}, suggestive of an interaction where radiative driving by the AGN is carving out significant portions of the molecular gas reservoir. 

Comparing the position of the radio emission knots to the [\ion{O}{3}] NLR, we see that the radio is intertwined with the ionized gas, as observed in other nearby radio-quiet AGN. However, as seen in Figure~\ref{fig:vectors}, if the peak of the most extended radio structure is used as an axis of feedback, the nuclear radio emission is not evenly distributed. Specifically, north of the nucleus, we see knots NE and C are located along the right side of the NLR against the western CO reservoir and not the left, eastern reservoir. We do see that extended radio emission to the south of the nucleus is located along the left side of the NLR, against the eastern CO reservoir, south of the nucleus. Separate from the VLBA detected knots, there is also a radio knot in Figure~\ref{fig:contours} south of the nucleus that also overlaps with [\ion{O}{3}] emission and is adjacent to CO emitting gas on the right, western side of the reservoir. 

Comparing these structures to the near-IR [\ion{Fe}{2}] emission, we see a conical morphology similar to that of the [\ion{O}{3}] emission to the north of the AGN, which originates from between the two CO reservoirs east and west of the nucleus. The flux distribution of the [\ion{Fe}{2}] is closer to that of the radio continuum emission, with the brightest [\ion{Fe}{2}] knot aligned with the right, western [\ion{O}{3}] emission knots. Further, south of the AGN, the [\ion{Fe}{2}] emission also wraps around the left, eastern CO reservoir, similar to the observed [\ion{O}{3}] and radio-continuum morphologies. Finally, we note that the flux distributions across wavebands are fairly consistent; brighter emission in the north and fainter emission to the south, which is consistent with all of the observed features being related and adjacent to one another.

In concert, this intertwined emission across the electromagnetic spectrum suggests a scenario where the observed radio knots in NGC~1068 form within the optical NLR, as traced by the observed [\ion{O}{3}] emission, which represents the intersection between AGN ionization and host galaxy disk material. This analysis supports the scenario where the observed radio emission is therefore a byproduct that occurs only at locations where winds are producing significant shocks in the radio-quiet AGN host galaxy, similar to what is seen in supernova remnants \citep{Zak14,Fis19b,Smi20,Fis21,Gli22}. Analysis of high-frequency continuum emission mapped by ALMA of these knots finds their spectral indexes to be negative ($\alpha \leq-1$) throughout, as expected for synchrotron-like emission \citep{Gar19}. From our multi-band comparisons in Figure~\ref{fig:contours}, we see that the distribution of the brightest synchrotron radio continuum is not filling in the entirety of the NLR and is concentrated on the western side to the north of the AGN and eastern side to the south. This is also where we see the most prominent [\ion{Fe}{2}] emission, which is an excellent diagnostic for shock-induced ionization  \citep{Stu02,Ina13}, and is frequently observed to be co-located with radio continuum emission in radio-quiet AGN \citep{Sto10,Rif11,Rif13,Rif15}. While the direct relationship is not clearly understood, shocks may create significant amounts of localized, low-energy ionizing radiation that permeates deeper into gas reservoirs than photons with ionization potentials greater than 13.6~eV, which are absorbed by hydrogen at shallower depths. 

In NGC~1068, the radio-continuum emission and [\ion{Fe}{2}] flux distribution, and thus shock distribution, are likely uneven across the NLR because these structures are formed in the densest gas reservoirs that are often observed entering the radiation field of the AGN. Analysis of the rotating molecular gas by \citealt{Gar19} shows that the molecular gas reservoir is rotating counterclockwise, such that the right, western reservoir north of the AGN and the left, eastern reservoir south of the AGN are rotating into the AGN radiation field. Therefore, the regions where we observe radio-continuum and [\ion{Fe}{2}] emission in NGC~1068 are likely often the face of the relatively pristine molecular gas rotating into the AGN radiation field and being impacted by AGN winds. 

The nearby AGN in NGC~4151 exhibits a similar case of gas reservoirs rotating into AGN radiation fields \citep{Sto10}, with H$_2$ 2.12\,$\mu$m lanes rotating clockwise into the observed AGN radiation fields and producing [\ion{Fe}{2}] and radio-continuum emission along the impacted face of the gas lane. Relationships between [\ion{Fe}{2}] and radio-continuum observed in NGC~1068, NGC~4151, and others suggest that the radio structures are often aligned with host disk material, or significant gas lanes, out to distances of $\sim$1000 pc. As another comparison with radio jet structures, jets do not exhibit any such preference in orientation and are not required to be adjacent to or intertwined with host disk material. Therefore, if the radio emission in NGC~1068 and other radio-quiet AGN were due to a radio jet, it is highly coincidental that they are continuously adjacent to line-emission structures which aligns their position to the plane of the host galaxy or significant dust lanes. 



\subsection{Extended Radio Emission in NGC~1068}

As both epochs of VLBA observation were focused on the nuclear regions of NGC~1068, analysis of interactions for the most extended radio emission, particularly the pointed, lobe-like structure to the northeast of the nucleus visible in Figure~\ref{fig:vlba_comp}, are outside the scope of this work. However, we suggest that the same underlying physics responsible for the nuclear radio structure produces these features as well. 

Kinematic analysis of [\ion{O}{3}] across the inner arcminute of NGC~1068 with VLT/MUSE \citep{Ven21} shows that the line width of AGN-ionized [\ion{O}{3}] gas is large (W$_{70} \sim$ 1000 km s$^{-1}$) at small radii and drops significantly (W$_{70} \sim$ 150 km s$^{-1}$) at the base of the lobe-like structure to larger radii. However, the [\ion{O}{3}] emission morphology is continuous in intensity between the nucleus and the narrow-line emission to the northeast. This discrepancy between kinematics and morphology does not suggest a jet interaction, as we would expect to see a peak in FWHM at the apex of the jet lobe. Instead, we suggest that the more extended radio emission resides in a shielded or offset region separate from the inner, turbulent emission region. We hypothesize that the radius where we observe the difference in line width represents the inner radius of a spiral arm that is being penetrated by high-velocity outflows. These winds are launched from smaller radii, which we observe in the regions sampled by our VLBA observations, and travel radially into the host lane material, ionizing and excavating it in the process (i.e. \cite{Mee23}). Material is not launched at angles that reach above the spiral arm lane. Therefore, the ionized gas we observe at the larger distances is not disturbed and remains in rotation, producing the observed narrow emission lines. As the excavation site is a cavity, the optical ionized gas counterpart of this process is obscured by the overlying dust lane along our line of sight, preventing us from observing the correspondingly large optical line-width emission. These winds also shock the interior surface of the cavity, forming the pointed, lobe-like structure we observe. 

Further analysis is required to examine these hypotheses. Specifically, near- and mid-infrared observations of the extended radio structures using integral field spectroscopy instruments such as JWST/NIRSpec, JWST/MIRI, and Gemini/NIFS would test for the presence of shocks and warm molecular gas adjacent to the edges of the radio emission as we see near the nucleus.

\subsection{Radio Loudness at High Redshifts}


\cite{Fis21} found that radio sources observed with both VLA and VLBA exhibited discrepancies between peak flux values, suggesting that point sources observed with larger beams are hosting a significant accumulation of extranuclear emission at small radii, potentially due to shock processes similar to what we suggest is occurring in NGC~1068. Unresolved radio sources at high redshift may be exhibiting similar processes, affecting our understanding of distant AGN with compact radio sources in two ways. First, AGN classified as radio-loud may actually be radio-quiet. The hardness ratio used in this classification, $R_{X}$, compares the X-ray and radio emission originating in the central engine. AGN emitting as a radio point source, with current resolution capabilities, may be interacting with their host material, producing shocks which in turn produce extranuclear radio emission unrelated to the central engine. Second, variability due to environmental interactions appears to occur on observable timescales. Knot~S2 observed in Epoch~1 of our VLBA observations either disappears or is replaced by a radially interior radio knot, suggesting that structures can move in to, or out of, the ionizing radiation field and affect the integrated radio flux of the circumnuclear region. At large redshifts, this change in structure morphology would be lost within the instrument beam, and appear as an increase or drop in flux. Alternatively, these variations in morphology may also affect the astrometric position of quasar radio sources with new peak flux positions rising over time.

We suggest that further study is required on the incidence of significant [\ion{Fe}{2}] emission in nearby AGN with extended radio sources and higher redshift AGN with unresolved radio sources. Determining if a relationship or specific ratio exists in nearby AGN where interactions with their host disk occur would produce a useful diagnostic in determining the source of radio emission in unresolved sources at higher redshift.

\subsection{Further Impacts on Radio Survey Studies}

Often in recent studies of extended radio structures in radio-quiet AGN, measurements are performed to quantify the amount of mechanical feedback that is injected into the host environment (i.e., \citealt{Jar19,Vil21,Gir22}). Findings from these studies suggest that mechanical feedback is not efficient in comparison to radiative-mode feedback. However, these radio signatures do trace the impact of the most significant winds on the host galaxy. Further, these locations represent the terminal radii where AGN winds are no longer able to drive material away from the central engine due to obstruction via dense host material and may be compressing gas in these regions, promoting an increase in star formation. As such, surveying the radio morphologies of radio-quiet AGN in search of extended structures can probe the incidence of compacting, positive feedback. Developing a census of how often AGN are significantly interacting with their hosts and the size of said interactions would then be available for inclusion in AGN feedback models.

\section{Conclusions} \label{section: Conclusions}

We compared VLBA 5 GHz radio observations taken 22 years apart for radio-quiet Seyfert 2 AGN NGC~1068 and measured the projected distance between two extranuclear flux peaks, knots NE and C, surrounding the central engine. Our main conclusions are as follows:

\begin{enumerate}
    \item Knots NE and C are well defined across both epochs, exhibit projected distance offsets $<$ 0.5~pc, and travel in directions perpendicular to the general NLR structure. Across the timescale between the two observation epochs, the measured velocities translate to non-relativistic velocities that are a few percent the speed of light and would require paths along our line of sight to reach significant fractions of the speed of light.
    
    \item The location of the nuclear radio continuum is shown to have an intricate connection with surrounding molecular and ionized gas, including co-location with molecular gas reservoirs that are rotating into the NLR ionization field and low-ionization [\ion{Fe}{2}] emission.
    
    \item The measured velocities and trajectories of the observed radio flux peaks do not perform as expected if the emission originated from a relativistic radio jet. We instead suggest that the flux peaks are not individual, jetted knots of radio plasma, but peak emission of larger structures whose locations change with host gas reservoir morphologies over observable time periods.
    
    \item These findings continue to be consistent with the explanation that extended radio structures observed in radio-quiet AGN are not jets, but byproducts formed via localized shocks, which in turn produce the observed synchrotron radiation at these locations.
    
\end{enumerate}

\acknowledgments
The authors thank the anonymous referee for their helpful comments that improved the clarity of this work. TCF is thankful for the support of the European Space Agency (ESA). The authors thank Zo Chapman, Julia Falcone, Enrique Lopez Rodriguez, Beena Meena, Garrett Polack, Mitchell Revalski, Maura Shea, and Krista Lynne Smith for helpful discussions while developing this manuscript. TCF thanks Jack Gallimore for providing the Epoch 1 VLBA observations and Rogemar Riffel for providing the Gemini/NIFS observations used in this work.

\vspace{5mm}
\facilities{VLBA, VLA, EVLA, ALMA, Gemini, HST; The HST data presented in this paper were obtained from the Mikulski Archive for Space Telescopes (MAST) at the Space Telescope Science Institute. The specific observations analyzed can be accessed via \dataset[DOI: 10.17909/nkex-gm67] {https://doi.org/10.17909/nkex-gm67}. }

\software{\textsc{aips}~\citep{van96}, Astropy~\citep{2013A&A...558A..33A}, \textsc{casa}~\citep{CASA22}
          }

\bibliography{ngc1068_vlba}

\appendix

\section{HST [\ion{O}{3}] Alignment} \label{section: alignment}

To investigate the relationship between the AGN in NGC~1068 and its surrounding environment in exquisite high spatial resolution, we utilized a myriad of archival data spanning a wide range of wavelengths across the electromagnetic spectrum, each of which probe various aspects of the AGN and its feedback on its surrounding medium. In this work, we used Hubble Space Telescope optical [\ion{O}{3}] narrow line emission as part of a suite of emission structures to suss out the source of the observed offsets of the peak emission features seen on the highest resolution VLBA spatial scales.  In this section, we describe our treatment of the data to account for any astrometry offsets that might exist between the various datasets.

\subsection{HST [\ion{O}{3}] Data}\label{subsection: HST data}

We obtained an [\ion{O}{3}] emission line image from the $\emph{HST}$ data archive as well as a broad spectrum continuum emission image that we used to subtract continuum emission sources from the [\ion{O}{3}] image, thus leaving only the narrow band [\ion{O}{3}] line emission in the central region of NGC~1068.  However, before applying the continuum subtraction, we used the unsubtracted [\ion{O}{3}] image to astrometrically align the continuum point sources in the central region of the galaxy to Gaia EDR3.  Once the astrometry was corrected, we then subtracted off the continuum emission from the [\ion{O}{3}] image and applied the astrometric solution to the resulting narrow band [\ion{O}{3}] emission line image.

\subsubsection{HST Astrometric Alignment} \label{subsubsection: hst astrometric alignment}

The native World Coordinate System (WCS) solution of the $\emph{HST}$ data is not sufficient for comparison with the high angular resolution and astrometric precision of the VLA/VLBA data, with astrometric offsets of 2--3 pixels (100--150~mas) visible by eye. We corrected the WCS solution of the $\emph{HST}$ data using the following procedure:
\begin{enumerate}
\item Query the Gaia~EDR3 catalog for all sources overlapping with the $\emph{HST}$ image and plot them as region files.
\item For Gaia sources associated with a compact knot of emission, manually center the region file on the knot. Delete remaining Gaia sources associated with extended emission. This produced 46 Gaia sources with compact knots in the $\emph{HST}$ image. \label{item: gaia source step}
\item Fit two-dimensional Gaussians to the Gaia associations to determine the precise centroid coordinates $(x,y)$ and convert to $(\alpha,\delta)$ using the original WCS solution. \label{item: xy}
\item Perform a least-squares 6-parameter fit (translation, rotation, scale) to the Gaia coordinates using the set of $(\alpha,\delta)$ from (\ref{item: xy}). To mitigate the effect of astrometric outliers, the 6-parameter fit was carried out in an iterative fashion, at each iteration using a random subset (without replacement) of the 46 Gaia sources from (\ref{item: gaia source step}). We used subsets of size 16, which was chosen as a balance between maximizing the degrees of freedom of the fit and minimizing outliers. New fit coordinates $(\alpha, \delta)$ are calculated at each iteration, as well as the astrometric residuals $\Delta\alpha = (\alpha-\alpha_\mathrm{Gaia})\cos{\delta}$, $\Delta\delta = \delta-\delta_\mathrm{Gaia}$. If the absolute mean values of $\Delta\alpha$, $\Delta\delta$ are less than their standard errors and the maximum value of the total offset $\sqrt{(\Delta\alpha)^2 + (\Delta\delta)^2}$ is less than that of the previous best WCS solution, the WCS solution of the iteration is stored as the new best. \label{item: iteration}
\end{enumerate}

\begin{figure}
\centering
\includegraphics[width=1\columnwidth]{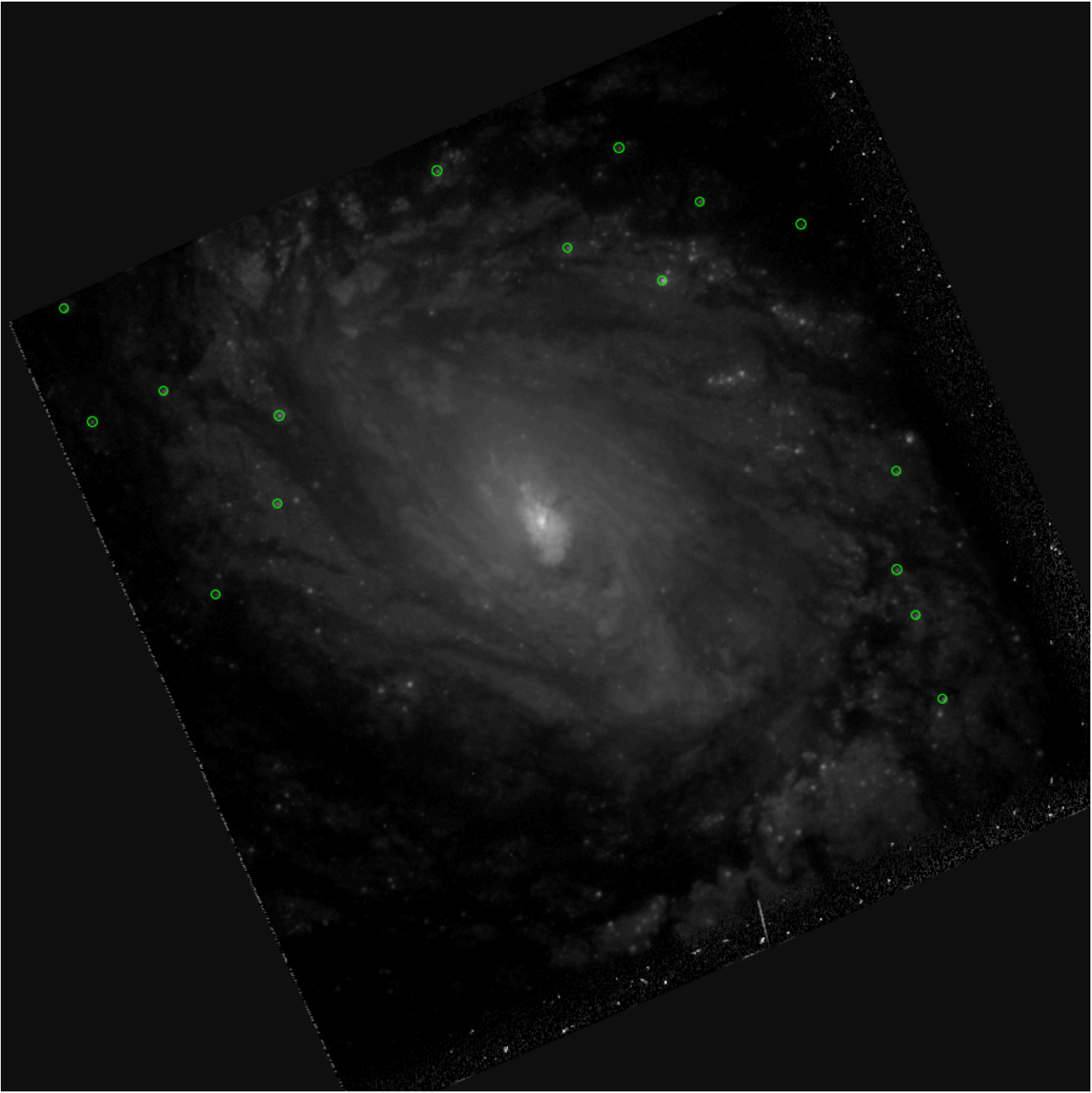}
\caption{Locations of Gaia~EDR3 astrometric reference points used to correct the $\emph{HST}$ astrometry, overlaid on the F547M image used for the counterparts.}
\label{fig:n1068_refs}
\end{figure}

\noindent By using the iterative fitting strategy in (\ref{item: iteration}), We were able to achieve high astrometric precision relative to Gaia~EDR3, with a mean offset in $\alpha$ and $\delta$ of $-0.6\pm1.5$~mas and $-0.9\pm1.7$~mas, where the uncertainties are the standard errors of the means. These 16 sources circumscribe NGC~1068 (Figure~\ref{fig:n1068_refs}) so the nuclear region is not affected by errors of extrapolation, and the 6-parameter fit does not include any tilt or higher-order distortion terms, which can introduce errors of interpolation.

\end{document}